\documentclass{WileyMSP-template}

\begin{document}

\pagestyle{fancy}
\rhead{\includegraphics[width=2.5cm]{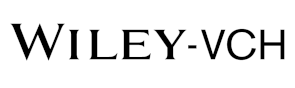}}

\title{Methodology for Topological Interface Engineering in 2D \\ Photonic Crystals}

\maketitle


\author{Ondřej Novák}
\author{Martin Veis}
\author{Gervasi Herranz*}

\dedication{}

\begin{affiliations}
Ondřej Novák, Martin Veis\\
Institute of Physics at Charles University\\
Ke Karlovu 2026/5, 121 16 Prague, Czech Republic\\
Email: \texttt{ondrej.novak@matfyz.cuni.cz}, \texttt{martin.veis@matfyz.cuni.cz}

Gervasi Herranz\\
Institut de Ciencia de Materials de Barcelona (ICMAB-CSIC)\\
Campus UAB, Bellaterra, 08193 Catalonia, Spain\\
Email: \texttt{gherranz@icmab.es}
\end{affiliations}



\begin{abstract}
Topological photonics provides a robust and flexible platform for controlling light, enabling functionalities such as backscattering-immune edge transport and slow-light propagation. In this work, we design and characterize photonic topological interfaces in two-dimensional photonic crystals. We introduce an iterative band connection algorithm that preserves mode symmetry and present a general framework for band symmetry recognition, essential for identifying $\mathbb{Z}_2$ topological phases. Design strategies for unit cell geometries are developed to achieve targeted band inversions, overlapping bandgaps, and tailored dispersions. Furthermore, the approach can be readily adapted to specific material platforms and operating wavelengths, including the telecommunication range, by appropriately scaling the lattice parameter as long as absorption remains low. We investigate the trade-off between bandgap size and band flatness, identifying exceptions governed by lattice geometry. Additionally, we demonstrate how photonic crystal periodicity influences the stability of topological modes and enhances unidirectional energy transport.

\end{abstract}


\section{Introduction}

Photonic crystals have emerged as a versatile platform for controlling light propagation through periodic variations in refractive index \cite{jones2013theory,vaidya2023topological}. Beyond traditional applications in guiding, filtering, and localizing light, the exploration of topological phenomena in photonic systems has opened new avenues for robust optical transport and novel device functionalities \cite{tkachov2015topological, kang2023topological, smirnova2020nonlinear, wang2022short,wang2008reflection}. Inspired by topological insulators in condensed matter physics, photonic topological systems provide a framework for designing light modes that are protected against scattering from disorder or imperfections, making them promising candidates for advanced photonic technologies \cite{bliokh2015quantum,liu2020z2,tang2022topological,kagami2020topological,zhang2023experimental}.

In two-dimensional (2D) photonic crystals \cite{khanikaev2017two}, the topology of photonic bands can be characterized using topological invariants such as the Chern number \cite{berry1984quantal,wang2008reflection,blanco2020tutorial}
or \(\mathbb{Z}_2\) indices \cite{xu2016accidental,wang2019band,huang2022topological}. These invariants arise from the symmetry properties of the photonic crystal and govern the existence of protected edge states \cite{haldane1988model}. Photonic analogs of quantum Hall and quantum spin Hall effects have been demonstrated, where carefully engineered photonic structures exhibit robust edge states immune to backscattering \cite{arora2022breakdown,li2021experimental,gong2020topological}. Designing and characterizing topological systems is challenging, with only a few known designs currently achieving topological protection. The pattern of the unit cell often appears incidental in producing topologically non-trivial behavior, making systematic design difficult \cite{xu2016accidental, huang2022topological, christiansen2019designing, christiansen2019topological}.

One major challenge is recognizing band symmetry and ensuring proper band connections. The symmetry of Bloch functions determines the topological state of the system. As the geometrical parameters of the unit cell are adjusted, the Bloch functions evolve, complicating the process of matching symmetry classes. When calculating the band structure, eigenproblems are solved at discrete \textbf{\textit{k}} points. To identify crossings and anti-crossings—key features for determining the Chern number—the bands must be connected based on the symmetry of their eigenfunctions \cite{blanco2020tutorial, devescovi2024tutorial}. Furthermore, for $\mathbb{Z}_2$ topology, the design revolves only around $C_6$ or honeycomb lattices and the scheme to achieve it for $C_4$ lattice is still missing or does not produce sufficient transmission or requires combination of multiple materials, which is inconvenient for the fabrication \cite{wu2015scheme,jin2022manipulation,lu2014topological,he2024wavelength,palmer2021berry,zhang2024topological,he2022topological,xiong2022topological}. Additionally, there is a lack of detailed information on characterizing topologically protected modes, particularly regarding the number of unit cells required to form the topological interface. This parameter is crucial for integrating topological photonics (toponics) into practical systems.

This paper focuses on addressing these challenges by presenting a systematic approach to the design and characterization of photonic topological interfaces in 2D photonic crystals. We analyze the eigenvalue problem derived from Maxwell's equations, propose an iterative method for band connection that preserves mode symmetry, and introduce a framework for symmetry recognition to classify bands for \(\mathbb{Z}_2\) topology. Furthermore, we demonstrate how unit cell designs can be tailored to achieve desired topological properties, such as gap overlaps and mode dispersion.

\section{Photonic topology in 2D}
The topological state of a system in 2D photonic crystal can be characterized by the topological invariants, which are derived from Maxwell's equations \cite{watanabe2019proof, blanco2020tutorial}. Suppose we have a photonic crystal with a unit cell size of $a$ and the material in the unit cell is a linear material with low dispersion and negligible absorption in the spectral region of interest \cite{newnham2005properties}. 
In two dimensions, the Maxwell equations decouples into transversal electric (TE) and transversal magnetic (TM) modes \cite{jones2013theory}. In this work, we will focus on the properties of TM modes. Isolating $E_z$ from the decoupled Maxwell equations, we obtain:

\begin{equation}
\nabla \times \nabla \times \mathbf{E}_z (\mathbf{r}) = \left( \frac{\omega}{c} \right)^2 \epsilon(\mathbf{r}) \mathbf{E}_z (\mathbf{r}),
\end{equation}

which can be written in a form of eigenproblem:

\begin{equation}
\hat{A} \textbf{E}_z (\mathbf{r}) = \hat{a} \textbf{E}_z (\mathbf{r}).
\label{eigenproblem}
\end{equation}

$\textbf{E}_z$ is a z-component of an electric field, $\textbf{r}$ is a spatial coordinate in the elementary cell, $\omega$ is an angular frequency of a wave, $c$ is a speed of light in vacuum, $\varepsilon(\textbf{r})$ is a spatially distribution of permittivity inside the photonic crystal elementary cell, $\hat{A}$ is an operator operating on eigenvector $\textbf{E}_z$, $\hat{a}$ is an eigenvalue of operator $\hat{A}$ with eigenvector $\textbf{E}_z$ and it carries meaning of reduced energy $\frac{a}{\lambda}$ or $\frac{\omega a}{2 \pi c}$. Since the $\varepsilon(\textbf{r})$ is periodical on $a$ and this $\hat{A}$ is also periodical, we can use the formalism of quantum mechanics and expand the eigenvectors into Bloch form:

\begin{equation}
    \label{bloch}
    E_{z,k} (\mathbf{r}) = e^{i \mathbf{k} \cdot \mathbf{r}} u_k (\mathbf{r}),
\end{equation}

where $u_k (\mathbf{r})$ is a periodical function on the elementary cell and will be the focus of this work.

\subsection{Photonic Chern topology}
Photonic Chern topology is a photonic analog to Quantum Hall effect, where the time reversal symmetry is broken by an external magnetic field \cite{wang2008reflection,wang2020universal}. The topological character of a photonic band is determined by its Chern number $C_n$, where $n$ indexes the band. The Chern number is defined as an integral of the Berry curvature $\Omega_n(\mathbf{k})$ over the Brillouin zone:

\begin{equation}
    C_n = \frac{1}{2\pi} \int_{\text{BZ}} \Omega_n(\mathbf{k}) \, d\mathbf{k}.
\end{equation}

This integral can be evaluated numerically using a finite grid in $\mathbf{k}$-space~\cite{blanco2020tutorial}, where the Berry curvature is expressed as the curl of the Berry connection:

\begin{equation}
    \Omega_n(\mathbf{k}) = \nabla_{\mathbf{k}} \times \mathbf{A}_n(\mathbf{k}).
\end{equation}

The Berry connection $\mathbf{A}_n(\mathbf{k})$ is defined as

\begin{equation}
    A_{mn}(\mathbf{k}) = i \bigl\langle u_{\mathbf{k},m} \big| \nabla_{\mathbf{k}} \big| u_{\mathbf{k},n} \bigr\rangle,
\end{equation}

where $|u_{\mathbf{k},n}\rangle$ are the periodic parts of the Bloch functions evaluated on a discretized $\mathbf{k}$-grid, and the indices $m, n$ refer to neighboring $\mathbf{k}$-points used to approximate derivatives. This formulation is valid in regions where the bands are non-degenerate and well-separated. However, in the presence of band crossings or near-degeneracies, this approach breaks down. In such cases, the Berry connection must be generalized using an overlap matrix that accounts for the mixing between all involved bands, as described in Refs.~\cite{blanco2020tutorial, marzari1997maximally}. A critical challenge arises in properly identifying band crossings. Numerical solvers such as MPB or COMSOL~\cite{johnson2001block} typically solve the eigenproblem (Eq.~\ref{eigenproblem}) and return eigenvalues ordered by magnitude, without regard to the physical symmetry of the associated eigenmodes. This can lead to incorrect band connections, especially near degeneracies, since eigenvalues alone do not uniquely identify bands. To address this, eigenvalues must be connected based on the symmetry of the corresponding eigenvectors. Our approach for robust band connection using eigenvector symmetry is detailed in Section~\ref{band connection}.

\subsection{Photonic $\mathbb{Z}_2$ topology}

The $\mathbb{Z}_2$ topological classification underpins the quantum spin Hall effect~\cite{bliokh2015quantum}. In photonic systems, a $\mathbb{Z}_2$ topological phase can arise when two gapped media with overlapping bandgaps are interfaced, provided there is an inversion in the symmetry ordering of the eigenmodes across the gap. For instance, in one system, photonic modes with $p$-like symmetry lie below the bandgap while $d$-like modes appear above it, whereas in the other system this ordering is reversed—the $d$-modes lie below the gap and the $p$-modes above. This symmetry inversion across the interface gives rise to topologically protected edge states.

To identify such an inversion, it is essential to recognize the symmetry of the photonic bands. However, this task is nontrivial, as the design of topological structures typically relies on heuristic searches through a high-dimensional geometric parameter space~\cite{christiansen2019designing, xu2016accidental, huang2022topological}. Consequently, even within the same symmetry group, the spatial distribution of the photonic fields can vary significantly, complicating symmetry identification. We address this challenge in Section~\ref{band recognition}.

\section{Band connection}
\label{band connection}

To understand and exploit the topological properties introduced in the previous section, it is necessary to reliably track the symmetry character of photonic bands across the Brillouin zone.
We first present a general method for identifying band connectivity across band crossings and regrouping bands according to their symmetry. The goal is to ensure that each reconstructed band maintains a consistent symmetry character along a path through the Brillouin zone. To achieve this, we connect bands such that the symmetry of the Bloch function $u_k(\mathbf{r})$ is preserved throughout the band. To do so, we define a distance between eigenvalues $\hat{a}_{n,k}$ and $\hat{a}_{m,k+1}$ (obtained by solving Equation~\ref{bloch}) as follows:
\begin{equation}
\label{distance_init}
    d_{m,n,k} = \frac{|\hat{a}_{n,k}-\hat{a}_{m,k+1}|}{\langle u_{n,k}(\mathbf{r}) |u_{m,k+1} (\mathbf{r}) \rangle},
\end{equation}

Here, $\hat{a}_{n,k}$ denotes the $n$-th eigenvalue obtained by solving Equation~\ref{eigenproblem} at a discrete index $k$ along the band structure path. Here $k$ is treated not as a wavevector, but as an index labeling points along a high-symmetry path $\zeta$ in the Brillouin zone. The proposed metric captures not only the energy differences, similar to the index-based matching approach used in MPB, but also the symmetry characteristics of the modes. This provides an efficient method for reconstructing the bands such that the symmetry of each band is preserved continuously along the path. Since the eigenmodes evolve smoothly with $k$, any deviation in symmetry between $u_{n,k}(\mathbf{r})$ and $u_{m,k+1}(\mathbf{r})$ results in a reduced scalar product, which in turn increases the distance defined in Equation~\ref{distance_init}, allowing reliable symmetry-based mode tracking.

To initialize the band reconstruction procedure, a suitable reference point in the Brillouin zone must be selected. Rather than starting at high-symmetry points such as $\Gamma$, where degeneracies are common, we choose a $k$-point, denoted $k_s$, that exhibits the largest energy separation between adjacent bands. This separation is quantified at each $k$ by

\begin{equation}
    S_k = \prod_{n=0}^{n_{\text{max}} - 1} \left( \hat{a}_{n+1,k} - \hat{a}_{n,k} \right),
\end{equation}

where $\hat{a}_{n,k}$ is the eigenvalue of the $n$-th band at momentum $k$. The point $k_s$ that maximizes $S_k$ is selected as the starting point for the distance-based band connection algorithm described in Eq.~\ref{distance_init}.

To systematically track band connectivity across the Brillouin zone, we introduce a new index $i$ to label the reconstructed bands, ensuring symmetry continuity along the path. The construction of the $i$-th band begins at $k_s$ with the eigenvalue $\hat{a}_{i,k_s}$. To extend this band to the next $k$-point, we identify the index $m$ corresponding to the eigenvalue $\hat{a}_{m,k}$ (identified among all the eigenvalues obtained by solving Equation~\ref{bloch}) that minimizes the distance metric $d_{m,i,k_s}$. The eigenvalue $\hat{a}_{m,k_{s+1}}$ is then assigned to the $i$-th band. Following this protocol, after traversing all $k$-points, we compute an averaged eigenvector for each reconstructed band to characterize its global symmetry:


\begin{equation}
    u_i(\textbf{r}) = \frac{1}{k_{max}+1}\sum_{k=0}^{k_{max}}u_{i,k}(\textbf{r})
\end{equation}

This averaged mode is then used in a modified distance metric $\Bar{d}$:

\begin{equation}
\label{distance}
    \Bar{d}_{i,m,k} = \frac{|\hat{a}_{i,k}-\hat{a}_{m,k+1}|}{\langle u_i(\mathbf{r}) |u_{m,k+1} (\mathbf{r}) \rangle},
\end{equation}

\begin{figure} [h]
    \centering
    \includegraphics[width=0.4\textwidth]{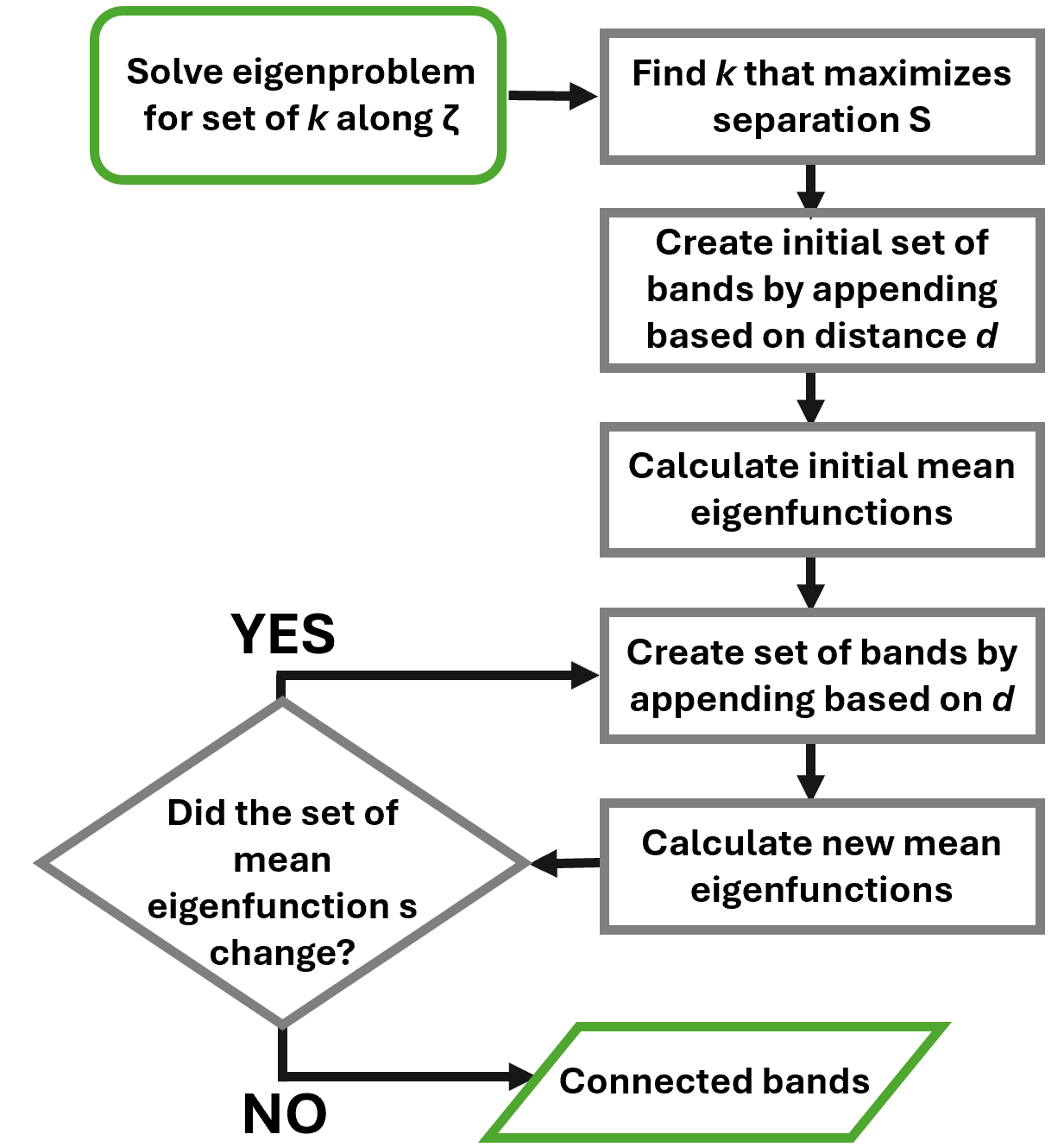}
    \caption{Flowchart of the band connection algorithm. The process begins by solving the eigenproblem along the $\zeta$ path, selecting the $\bm{k}$ point with the highest separation $S$, and constructing an initial set of bands based on the distance $d$. The mean eigenfunctions are computed iteratively, and bands are updated based on $\bar{d}$. The algorithm repeats until the mean eigenfunctions remain unchanged, indicating convergence and successful band connection.}
    \label{diagram}
\end{figure}

Note that in the denominator of $\Bar{d}_{i,m,k}$, the $m$-th eigenfunction is now compared to the average eigenfunction of the $i$-th band. This modified distance metric, denoted by $\Bar{d}$, is then used to reconstruct the bands. The process of eigenmode averaging and band appending is performed iteratively until all eigenfunctions $u_i(\mathbf{r})$ converge. Iteration is crucial because, near degeneracies, eigenfunctions corresponding to nearly identical eigenvalues can mix, making it difficult to assign them consistently to the correct band. If only adjacent eigenmodes are considered, the algorithm may incorrectly follow an unintended eigenvalue branch. This iterative scheme is inspired by clustering algorithms commonly employed in machine learning \cite{xu2005survey,ahuja2020classification}, where consistent grouping is achieved through repeated refinement. The flowchart of the algorithm is depicted in figure \ref{diagram}.

Once the bands are constructed, a crossing between the $i$-th and $j$-th bands can be detected by examining their eigenvalues across $k$-points. Specifically, a crossing is identified when the sign of the difference between $\hat{a}_{i,k}$ and $\hat{a}_{j,k}$ changes between adjacent $k$-points.

\section{Band symmetry recognition}
\label{band recognition}

Having outlined a general method for regrouping bands based on their symmetries along high-symmetry paths in the Brillouin zone, we now apply this approach to analyze the band symmetries of crystals exhibiting $\mathbb{Z}_2$ topology. Identifying band symmetries in topological systems is inherently more challenging, as it requires comparing field profiles at specific $\mathbf{k}$-points across multiple geometric configurations. This added complexity arises from the need to explore how varying the geometry of photonic crystals maps out regions of trivial and non-trivial topology in parameter space. This stands in contrast to the band connection procedure described in the previous section, which only involves matching eigenmodes at different $\mathbf{k}$-points within a single geometry.

More specifically, identifying band symmetries in systems with $\mathbb{Z}_2$ topology involves determining the values of the geometric parameters of the unit cell at which band symmetry inversion occurs. To do this, we begin with a collection of eigenmodes computed across various unit cell geometries, with the goal of identifying modes that exhibit the same symmetry characteristics. This enables us to group eigenmodes into distinct symmetry classes. To facilitate this classification, we leverage the orthogonality properties of eigenmodes. However, since these modes originate from different eigenproblems, each corresponding to a different unit cell geometry, their field distributions may vary, even when governed by the same underlying lattice symmetry group. To simplify notation, we denote eigenmodes from all geometries and bands with a single index \(i\), and omit explicit reference to $\mathbf{k}$, as all eigenmodes are evaluated at the same $\mathbf{k}$-point, typically the $\Gamma$ point. Thus, we write each eigenmode as \(u_i(\mathbf{r})\).

To reduce the influence of amplitude variations caused by differences in unit cell geometry, while retaining the essential symmetry information, we apply the sign function to each eigenmode:

\begin{equation}
    v_i(\mathbf{r}) = \operatorname{sign}(u_i(\mathbf{r})).
\end{equation}

This transformation preserves the spatial symmetry of the mode while discarding its amplitude profile. For symmetry recognition, we define a set of comparator functions \( C_n(\mathbf{r}) \) that reflect the symmetry operations of the underlying lattice. In cases where the unit cell parameters interpolate between different lattice types, resulting in changes to the symmetry group, the comparator functions cannot be predefined. Instead, they must be inferred from the dataset itself and subsequently labeled based on their symmetry properties. This procedure is discussed in detail in the following sections.

\subsection{Creating comparators based on lattice types}
For the sake of clarity, let us apply this mode to the specific case of the  \(C_4\) point symmetry. To distinguish between \(s\)- and \(p\)-like modes in a \(C_4\) lattice, we construct comparator functions based on the irreducible representations of the \(C_4\) point group. These functions reflect the symmetry properties of the eigenmodes and are defined as:

\[
C_{px}(\mathbf{r}) = \operatorname{sign}(x), \quad 
C_{py}(\mathbf{r}) = \operatorname{sign}(y),
\]
\[
C_{d_{xy}}(\mathbf{r}) = \operatorname{sign}(xy), \quad 
C_{d_{x^2 - y^2}}(\mathbf{r}) = \operatorname{sign}(x^2 - y^2),
\]

Each comparator corresponds to a distinct symmetry class: \( C_{px} \) and \( C_{py} \) capture the odd (dipolar) symmetries along the \(x\)- and \(y\)-axes, associated with \(p\)-like modes; \( C_{d_{xy}} \) and \( C_{d_{x^2 - y^2}} \) capture the quadrupolar symmetries typical of \(d\)-like modes. These symmetry patterns are linked to specific irreducible representations of the \(C_4\) group, as discussed in \cite{sakoda2005optical}. For each eigenfunction \(v_n(\textbf{r})\), each comparator is evaluated using the scalar product. If the highest scalar product value among the comparators $m$ exceeds \(\frac{2}{3}\) of the eigenfunction resolution ($r_e\times r_e$ pixels), it is considered reliable to assign the corresponding label $m_{max}$ to the eigenfunction.

\begin{equation}
    \text{if }\exists m: \sum_{i,j}C_m[i,j]\cdot v_n[i,j]>\frac{2}{3}r_e^2\Rightarrow v_{n,m_{max}}
\end{equation}

\subsection{Generating Comparator Functions from the Dataset}

As discussed above, varying the geometric parameters of photonic crystal unit cells can lead to changes in point-group symmetry and, consequently, in the relevant irreducible representations. In such cases, predefined comparator functions may become insufficient, and a more adaptive, data-driven approach is required. 

To address this, we propose a method that generates comparator functions directly from the dataset of eigenmodes. The procedure begins by initializing an empty set of comparators, denoted as $\mathbb{C}$. Each eigenmode $u_n(\mathbf{r})$ is first processed via a sign function to produce $v_n(\mathbf{r}) = \operatorname{sign}(u_n(\mathbf{r}))$, which captures spatial symmetry while eliminating amplitude variations. Next, $v_n$ is compared against all existing comparators in $\mathbb{C}$ using a scalar product over the spatial grid. If no scalar product exceeds a predefined threshold, taken as two-thirds of the eigenmode resolution $r_e^2$, then $v_n$ is considered to represent a new symmetry class and is appended to the comparator set. In this way, $\mathbb{C}$ is incrementally constructed to reflect the full diversity of symmetry in the data set. The condition for appending a new comparator can be formally expressed as:

\begin{equation}
    \text{if } \nexists\, C_m \in \mathbb{C} \text{ such that } \sum_{i,j} C_m[i,j] \cdot v_n[i,j] > \frac{2}{3} r_e^2, 
    \quad \text{then append } v_n \text{ to } \mathbb{C}.
\end{equation}

\begin{figure} [h]
    \centering
    \includegraphics[width=0.6\textwidth]{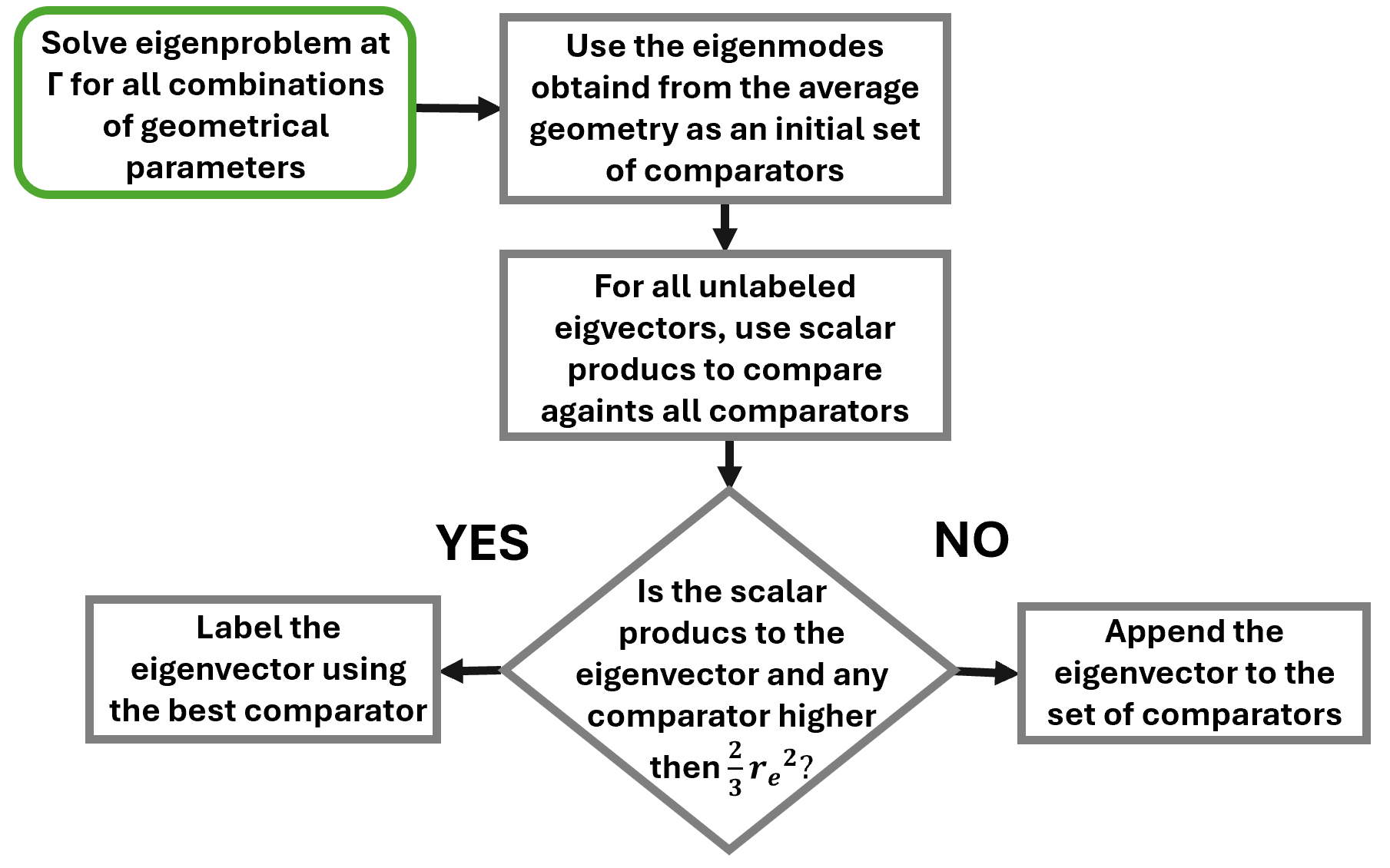}
    \caption{Flowchart illustrating the process of comparator generation and eigenvector classification. The eigenproblem is solved at the $\Gamma$ point for all combinations of geometric parameters. Eigenmodes from a representative geometry (typically near the center of the configuration space) are used to initialize the comparator set. Each unlabeled eigenmode is compared to all comparators via scalar products. If a match exceeding the threshold $\frac{2}{3} r_e^2$ is found, the eigenmode is assigned to the corresponding class. Otherwise, it is appended to the set of comparators.}
    \label{diagram_generation}
\end{figure}

An important consideration is that eigenmodes corresponding to extreme geometries, that is, those at the edges of the parameter space, may exhibit atypical features and introduce noise into the comparator set. To mitigate this, we recommend initiating the comparator generation process using eigenmodes from the central region of the parameter space, where the geometries are more representative. Once a robust set of comparators has formed from this core, the remaining eigenmodes, particularly those of the edge geometries, can be incrementally classified.
The flow chart illustrating the entire procedure for the generation of comparators and the classification of eigenvectors is shown in Figure~\ref{diagram_generation}.

\section{Unit cell design}

When designing a photonic crystal lattice, the expected eigenmode symmetries are typically known in advance, dictated by the point-group symmetry of the lattice. The permittivity distribution within the unit cell not only defines the spatial profiles and eigenvalues of the modes, but also governs the overall band ordering. To realize nontrivial \(\mathbb{Z}_2\) topological phases, a band inversion is often required, that is, a deliberate reversal in the energy ordering of modes with distinct symmetry characters. This can be achieved by tailoring the permittivity profile to enhance the spatial overlap with a specific mode, thereby lowering its eigenvalue and shifting it to lower energy.

\begin{figure} [h]
\centering
\includegraphics[width=0.6\textwidth]{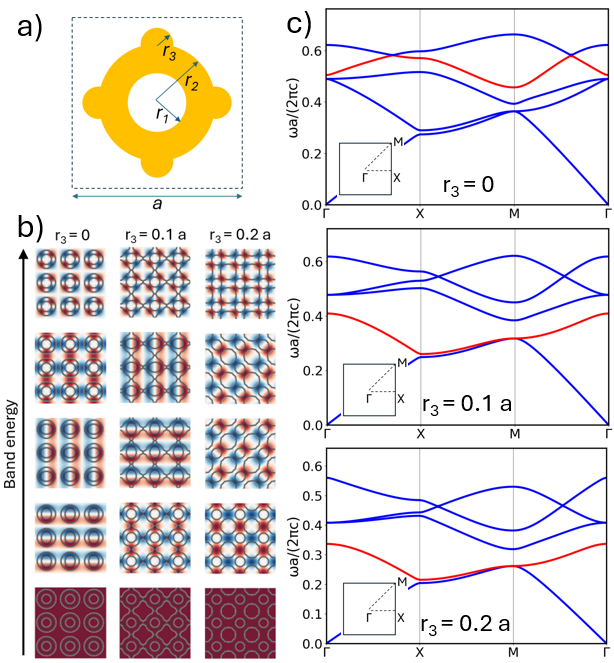}
\caption{\label{permitivity overlap}Evolution of Photonic Crystal Band Structure and Eigenmodes with Varying Mode-Permittivity Overlap.  
\textbf{a)} Definition of the 2D photonic crystal unit cell. The material used is indium phosphide (yellow). Fixed parameters: \( r_1 = 0.3a \), \( r_2 = 0.4a \).  
\textbf{b)} Eigenmodes for TM polarization at the \(\Gamma\)-point, computed from Equation~\ref{bloch}, $E_z$ shown. Columns correspond to increasing values of \( r_3 \); rows show different bands.  
\textbf{c)} Band structures for varying \( r_3 \). The red band highlights the \( d_{x^2 - y^2} \) mode, which shifts to lower energies as \( r_3 \) increases.
}
\end{figure}

Figure~\ref{permitivity overlap} illustrates this concept in a two-dimensional PC. The unit cell is defined by three geometric parameters: the inner ring radius \( r_1 \), the outer ring radius \( r_2 \), and the radius \( r_3 \) of four cylinders placed near the ring's edge. Indium phosphide (yellow) is used as the high-permittivity material, with fixed values \( r_1 = 0.3a \) and \( r_2 = 0.4a \), while \( r_3 \) is varied. Figure~\ref{permitivity overlap}\textbf{b} shows the evolution of transverse magnetic eigenmodes at the \(\Gamma\)-point as \( r_3 \) increases, while Figure~\ref{permitivity overlap}\textbf{c} displays the corresponding band structures. As \( r_3 \) increases, the overlap between the high-permittivity regions and the \( d_{x^2 - y^2} \) mode strengthens, lowering its eigenvalue and pushing it below the \( p_x \) and \( p_y \) modes. This targeted band inversion is a critical mechanism for achieving partial \(\mathbb{Z}_2\) topological protection~\cite{xu2016accidental}. While the primary effect is observed in the \( d_{x^2 - y^2} \) mode, other bands are also influenced by changes in \( r_3 \), due to the global modification of the permittivity distribution.

A key challenge in this design strategy lies in the need to explore a broad geometric parameter space to identify suitable unit cell configurations capable of producing full or partial band inversions. To create a functional topological interface, one must pair such a geometry with a second unit cell that does not exhibit inversion, and crucially, both configurations must support photonic bandgaps that overlap in frequency. This overlap is essential: it ensures that the topological edge states reside within a spectral range where neither bulk configuration supports propagating modes. Without such an overlap, although topological modes could still form at the interface, they would fall within the continuum of bulk modes, allowing them to scatter into the surrounding medium. This scattering undermines one of the key advantages of topological systems, their protection against backscattering. Thus, the presence and alignment of bandgaps in both structures are critical to maintaining robust, protected edge transport.

While the design space offers some flexibility, the specific choice of unit cells critically influences key properties such as the band-gap size, the slope of the topological dispersions, and whether the dispersion is normal or anomalous. These factors not only affect the existence and robustness of edge states but also determine how effectively they can be used in practical photonic applications.

\begin{figure} [h]
\centering
\includegraphics[width=0.6\textwidth]{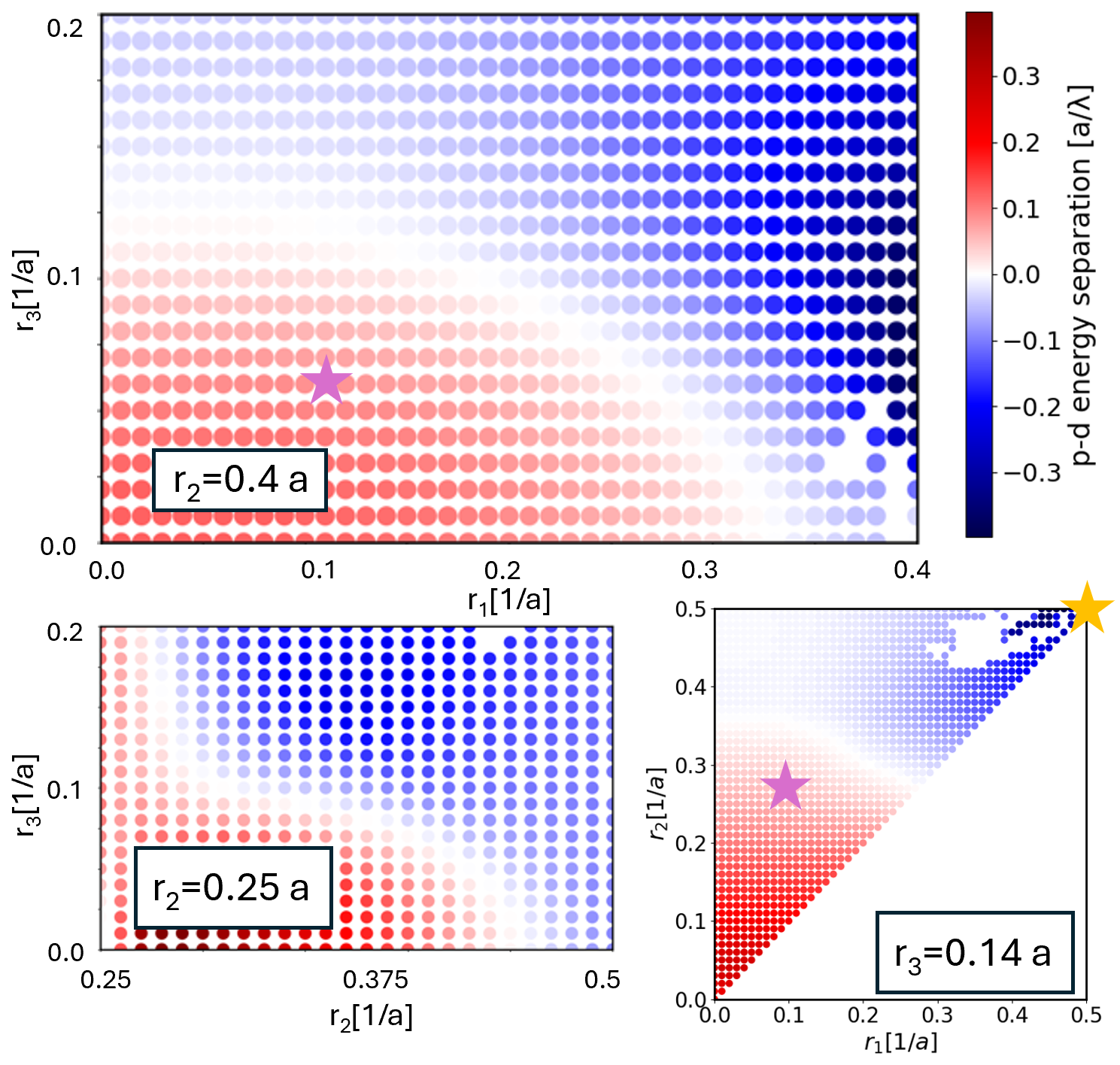}
\caption{\label{phase diagram}Phase diagram of partial \(\mathbb{Z}_2\) band inversion for the structure shown in Fig.~\ref{permitivity overlap}\textbf{a}. The color scale indicates the energy difference between the \( d_{x^2 - y^2} \) mode and the \( p \)-modes at the \(\Gamma\)-point. Positive values (red) correspond to configurations where the \( d \)-mode lies above the \( p \)-modes, indicating no band inversion. Negative values (blue) correspond to inverted configurations, where the \( d \)-mode lies below. Each subplot represents a 2D slice through the 3D configuration space defined by geometric parameters \( r_1 \), \( r_2 \), and \( r_3 \). Missing data points result from failure in automatic state identification due to near-degeneracies or ambiguous mode classification.}
\end{figure}

To evaluate the design flexibility offered by our symmetry-tracking framework, we examine the formation of topological interfaces in two representative scenarios using the photonic crystal structure defined in Fig.~\ref{permitivity overlap}. The central metric guiding this design process is the energy difference between the \( d_{x^2 - y^2} \) mode and the nearest \( p \)-mode at the \(\Gamma\)-point, as visualized in the phase diagram in Fig.~\ref{phase diagram}. This diagram maps the occurrence of band inversion across a broad range of geometric parameters, offering a practical guide for selecting inverted and regular unit cells.

In the first scenario, we select the configuration exhibiting the strongest band inversion, indicated by the yellow star in the \( r_1 \)-\( r_2 \) slice of Fig.~\ref{phase diagram}. This configuration corresponds to geometric parameters \( r_1 = 0.5a \), \( r_2 = 0.5a \), and \( r_3 = 0.14a \), where the \( d_{x^2 - y^2} \) mode is energetically well below the \( p \)-modes. To form a topological interface, we identify a complementary, non-inverted unit cell with maximal band-gap overlap. This configuration is marked by the purple stars in the same diagram and corresponds to \( r_1 = 0.1a \), \( r_2 = 0.275a \), and \( r_3 = 0.06a \). Two purple stars are shown to indicate the nearest match across adjacent slices in the 3D configuration space, as the exact coordinates do not lie on a single plane.

\begin{figure} [h]
\centering
\includegraphics[width=1\textwidth]{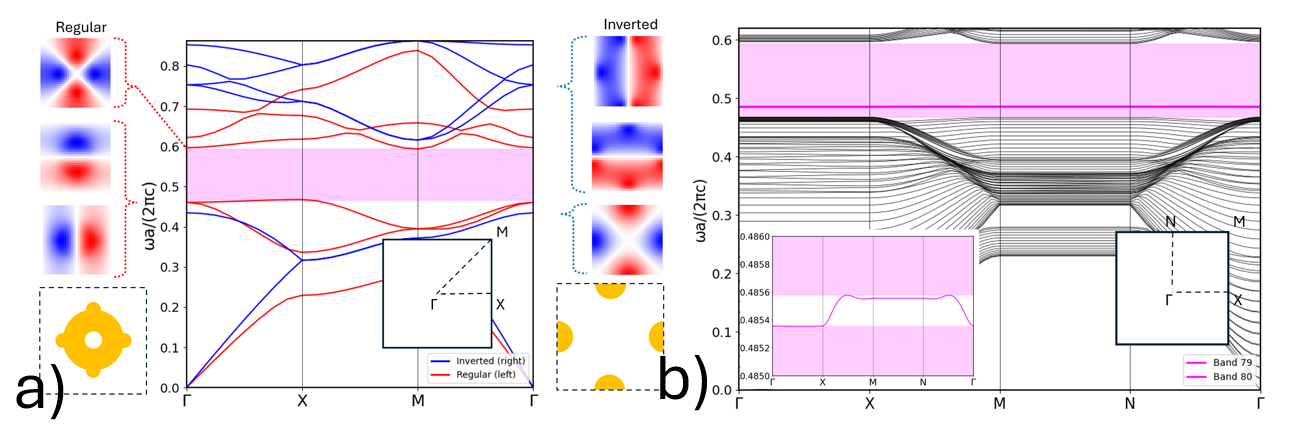}
\caption{\label{broad modes}\textbf{a)} Overlaid band structures of the photonic crystal defined in Fig.~\ref{permitivity overlap}\textbf{a}. The regular configuration (\( r_1 = 0.1a \), \( r_2 = 0.275a \), \( r_3 = 0.06a \)) and the inverted configuration (\( r_1 = 0.5a \), \( r_2 = 0.5a \), \( r_3 = 0.14a \)) are both composed of indium phosphide. \textbf{b)} Band structure of the interface formed by 16 alternating unit cells of the two designs. The magenta bands represent the topological interface modes with anomalous dispersion along both \( X\text{-}M \) and \( \Gamma\text{-}N \) directions.
}
\end{figure}

Figure~\ref{broad modes}\textbf{a} displays the overlaid band structures for both the regular and inverted configurations. The significant overlap between the bandgaps of the two unit cells ensures the formation of an interface that supports topological edge states. The interface band structure, formed by stitching together 16 unit cells of each design, is shown in Fig.~\ref{broad modes}\textbf{b}. The magenta-colored interface modes, localized at the domain wall, display anomalous (nonlinear and nonmonotonic) dispersion along both the \( X\text{-}M \) and \( \Gamma\text{-}N \) directions. These nearly flat bands indicate slow group velocity and are highly desirable for slow-light applications~\cite{arregui2021quantifying,kumar2024slow,xie2021topological}.

In the second scenario, we explore a more general use case, where the inverted configuration is chosen arbitrarily rather than optimized for maximal inversion. The selected parameters are \( r_1 = 0.345a \), \( r_2 = 0.425a \), and \( r_3 = 0.108a \). A regular configuration is then identified to provide optimal band-gap overlap with this inverted cell, with parameters \( r_1 = 0.105a \), \( r_2 = 0.35a \), and \( r_3 = 0.072a \).

\begin{figure} [h]
\centering
\includegraphics[width=1\textwidth]{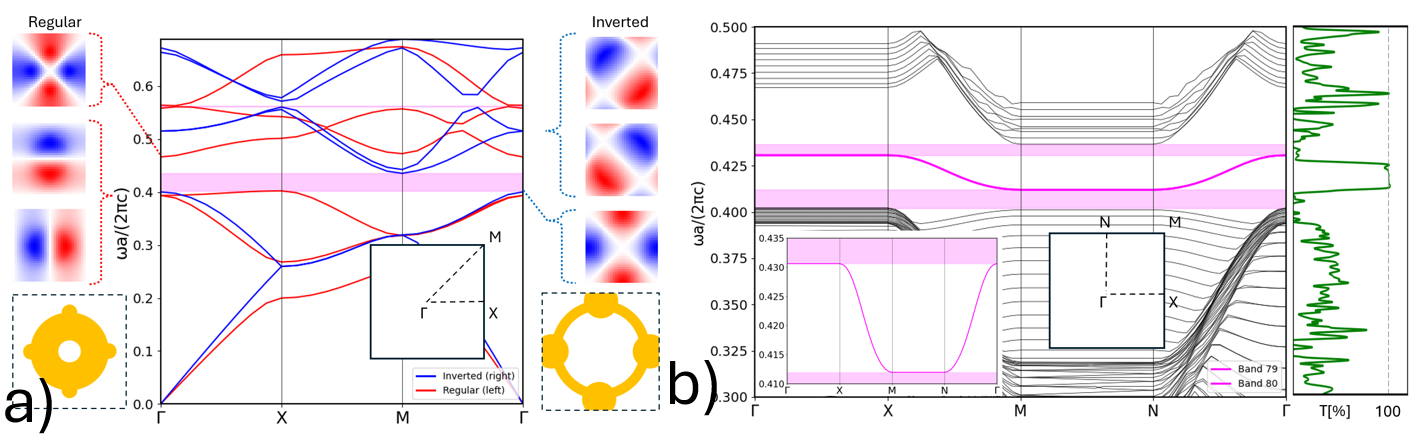}
\caption{\label{reduced modes}\textbf{a)} Overlaid band structures of the regular configuration (\( r_1 = 0.105a \), \( r_2 = 0.35a \), \( r_3 = 0.072a \)) and the inverted configuration (\( r_1 = 0.345a \), \( r_2 = 0.425a \), \( r_3 = 0.108a \)) for the structure defined in Fig.~\ref{permitivity overlap}\textbf{a}. \textbf{b)} Band structure of the interface formed by 16 unit cells of each configuration. The magenta bands denote the topological interface modes with regular dispersion. The green line shows the transmission spectrum computed using the FDTD method~\cite{sullivan2013electromagnetic,inan2011numerical,oskooi2010meep}, confirming efficient topological transport.
}
\end{figure}

Figure~\ref{reduced modes}\textbf{a} presents the overlaid band structures of these two configurations, highlighting the region of band-gap overlap. The corresponding interface band structure, built from 16 unit cells of each type, is shown in Fig.~\ref{reduced modes}\textbf{b}. In this case, the topological modes (again shown in magenta) exhibit regular, nearly linear dispersion, allowing clear definition of group velocity. The green curve overlays the interface band structure and represents the transmission spectrum calculated using the finite-difference time-domain (FDTD) method~\cite{sullivan2013electromagnetic,inan2011numerical,oskooi2010meep}, confirming the topological nature and high transmission of the edge modes.

\subsection{Band-Gap Scaling and Dispersion Characteristics}

In addition to symmetry and topological phase requirements, the dispersion characteristics of topological modes, particularly their slope, are crucial for applications involving energy transport and slow light. The gradient of the topological band determines the group velocity of the edge modes and directly impacts signal delay and light–matter interaction strength. Therefore, understanding how this gradient behaves as a function of geometric design and band-gap size is key to optimizing performance.

\begin{figure} [h]
\centering
\includegraphics[width=1\textwidth]{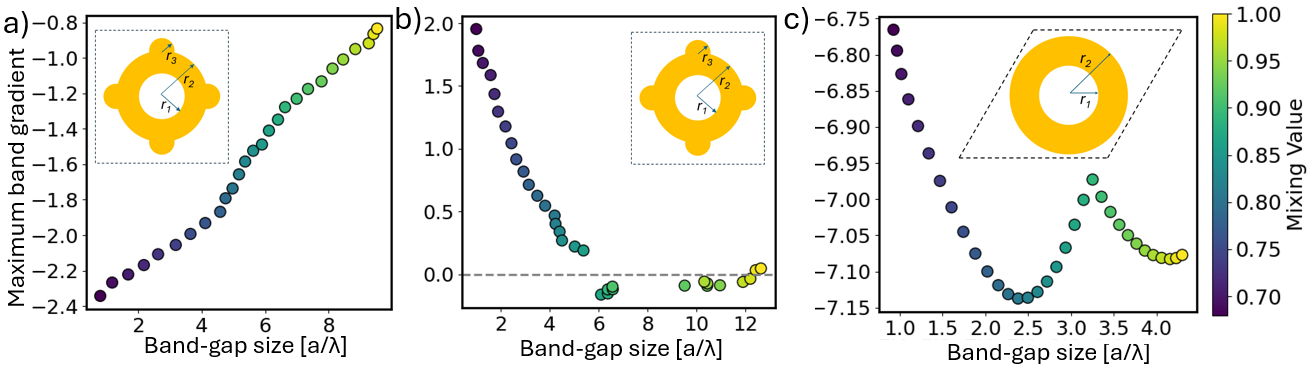}
\caption{\label{gap scaling}Dependence of the topological band’s maximum gradient along the \( X\text{-}M \) direction on the band-gap overlap of individual designs. All structures use indium phosphide (InP)~\cite{adachi1989optical} as the dielectric material. Intermediate geometries are generated by linear interpolation between regular and inverted designs using mixing values between 0.7 and 1.  
\textbf{a)} Interface formed from the structure defined in Fig.~\ref{permitivity overlap}, using: inverted (\( r_1 = 0.425a \), \( r_2 = 0.45a \), \( r_3 = 0.12a \)) and regular (\( r_1 = 0.025a \), \( r_2 = 0.325a \), \( r_3 = 0.06a \)). This pair yields the largest band-gap overlap.  
\textbf{b)} Interface formed from an alternative configuration where ring elements are replaced by cylinders for simplified fabrication. The inverted geometry is (\( r_1 = 0.5a \), \( r_2 = 0.5a \), \( r_3 = 0.14a \)), and the regular geometry is (\( r_1 = 0.1a \), \( r_2 = 0.275a \), \( r_3 = 0.06a \)).  
\textbf{c)} Interface derived from a hexagonal (\( C_6 \)) lattice structure [cite]. The scaling parameters are: inverted (\( r_1 = 0.38a \), \( r_2 = 0.45a \)) and regular (\( r_1 = 0.27a \), \( r_2 = 0.36a \)). Unlike \textbf{a} and \textbf{b}, this configuration exhibits non-monotonic behavior in band slope.
}
\end{figure}

Our analysis reveals a general trend: in certain configurations, particularly those involving partial band inversion, the gradient of the topological mode decreases as the band-gap overlap increases. In other words, achieving a wider band-gap often comes at the cost of slower mode propagation. This inverse relationship is clearly shown in Fig.~\ref{gap scaling}\textbf{a}, where the largest slope (gradient) of the topological band, computed along the \( X\text{-}M \) direction, is plotted against the magnitude of band-gap overlap between inverted and regular configurations. Here, negative gradient values correspond to normal (monotonic) dispersion, while a reduction in slope indicates flatter bands and slower group velocities. Interestingly, Fig.~\ref{gap scaling}\textbf{b} demonstrates that the size of the band-gap can also affect the nature of dispersion. In some geometries, dispersion transitions from anomalous (nonmonotonic) to normal as the gap increases. This transition is critical for selecting the appropriate design depending on whether one prioritizes dispersion linearity or group velocity reduction.

To generate the data shown in Figs.~\ref{gap scaling}\textbf{a} and \textbf{b}, we followed a systematic scaling procedure. Two seed geometries are selected from the configuration space (an inverted unit cell \( i_s \), and a regular (non-inverted) unit cell \( r_s \)) both of which are known to produce band-gap overlap at the interface. A series of intermediate geometries is then defined by interpolating linearly between these two points using a mixing parameter \( m \in [0, 1] \), as described by

\begin{equation}
\begin{aligned}
\label{mixing}
    i &= m i_s + (1 - m) r_s, \\
    r &= (1 - m) i_s + m r_s.
\end{aligned}
\end{equation}

For each value of \( m \), a corresponding interface is constructed, and the gradient of the resulting topological band is computed. Specifically, Fig.~\ref{gap scaling}\textbf{a} corresponds to an interface built from designs defined in Fig.~\ref{permitivity overlap}. The inverted geometry uses \( r_1 = 0.425a \), \( r_2 = 0.45a \), and \( r_3 = 0.12a \), while the regular geometry uses \( r_1 = 0.025a \), \( r_2 = 0.325a \), and \( r_3 = 0.06a \). This configuration yields the largest band-gap overlap in our dataset, providing a clear setting to analyze trade-offs between gap size and slope.

In Fig.~\ref{gap scaling}\textbf{b}, a different pair of seed geometries is used, also defined in Fig.~\ref{permitivity overlap}. The inverted structure (\( r_1 = 0.5a \), \( r_2 = 0.5a \), \( r_3 = 0.14a \)) features a simplified fabrication design where the ring structure is replaced with cylinders, and the regular geometry is \( r_1 = 0.1a \), \( r_2 = 0.275a \), \( r_3 = 0.06a \). This pair demonstrates a similar trend between band-gap size and dispersion steepness, although the trade-off appears somewhat relaxed compared to the example in \textbf{a}.

However, this inverse relationship does not hold universally. As shown in Fig.~\ref{gap scaling}\textbf{c}, for a design based on a different lattice geometry (more specifically, a pattern of rings arranged in a \( C_6 \) lattice) this behavior breaks down. In this case, the gradient of the topological modes exhibits local maxima as a function of the band-gap size, indicating a more complex dependence on lattice structure. The inverted and regular seed geometries used here are \( r_1 = 0.38a, r_2 = 0.45a \) and \( r_1 = 0.27a, r_2 = 0.36a \), respectively. This result underscores that band-gap scaling trends are highly sensitive to lattice symmetry and geometry, and that generalized design rules must be treated cautiously when transitioning between lattice types. 
To avoid undesirable trade-offs between band-gap size and dispersion, designers should begin by specifying the operational bandwidth required for their application, and then explore geometric configurations that yield both sufficient band-gap overlap and favorable dispersion within this region, while validating these trends across the relevant lattice symmetries.

\section{Design Characterization}

In addition to symmetry considerations, band inversion, and dispersion control, a crucial final design parameter is the number of photonic crystal periods surrounding the interface. This structural factor plays a central role in the emergence and robustness of topological modes, influencing their energy convergence, spatial localization, and transmission efficiency.

\begin{figure} [h]
\centering
\includegraphics[width=1\textwidth]{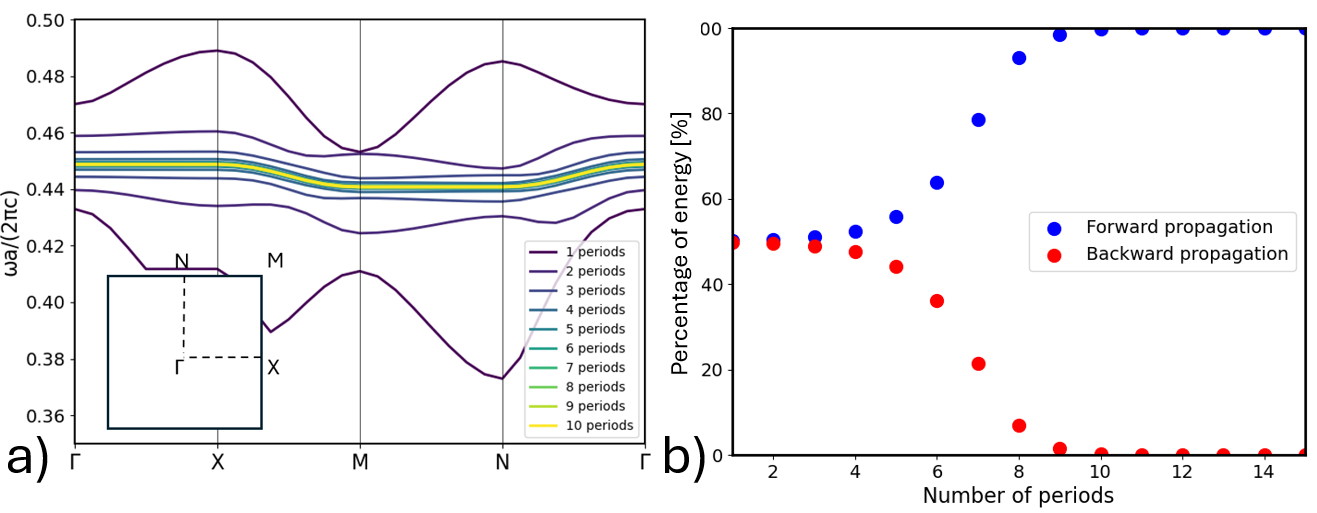}
\caption{\label{periods effect}Characterization of the photonic design defined in Fig.~\ref{reduced modes}.  
\textbf{a)} Evolution of the band structure as a function of the number of PC periods surrounding the interface. Convergence is achieved after approximately five periods, indicating stabilization of topological interface modes.  
\textbf{b)} Proportion of forward and backward energy transmission versus number of periods. At 10 periods, 99.7\% of the energy propagates forward, increasing to 99.99\% at 12 periods.
}
\end{figure}

To investigate this, we focus on the topological interface design introduced in Fig.~\ref{reduced modes}. Figure~\ref{periods effect}\textbf{a} shows the evolution of the band structure as the number of periods increases symmetrically around the interface. The data indicate that the energies of the topological modes converge rapidly, stabilizing after approximately five periods. This convergence ensures that the interface modes occupy well-defined energy levels, which is essential for applications requiring narrowband operation and frequency-selective guidance. In contrast, insufficient periods can lead to poorly localized or poorly defined modes, resulting in significant spectral broadening or even the loss of topological characteristics.

Beyond spectral stability, the number of periods also affects the degree of unidirectionality in energy propagation. In \(\mathbb{Z}_2\) topological photonics, where the system emulates the quantum spin-Hall effect~\cite{tkachov2015topological,bliokh2015quantum}, each topological mode is associated with a pseudospin corresponding to clockwise or counterclockwise circulation. For such unidirectional behavior to emerge, the interface must be formed by two semi-infinite PCs with different symmetry orderings and overlapping bandgaps. If the number of periods is insufficient, the system behaves more like a trivial defect or weakly coupled cavity, and fails to support genuine topological modes.

Figure~\ref{periods effect}\textbf{b} quantifies the impact of the number of periods on directionality. For the structure defined in Fig.~\ref{reduced modes}, we track the proportion of energy that propagates forward versus backward along the interface. At 10 periods, the forward-directed energy propagation reaches 99.7\%, indicating a well-defined topological mode with strong suppression of backscattering. Increasing the number of periods to 12 pushes this value even higher, reaching 99.99\%. This data clearly shows that increasing the number of interface-adjacent periods enhances the topological protection and reinforces the spatial separation of pseudospin channels.

To further examine how these modes localize within the interface region, we analyze Bloch mode profiles in supercells composed of both trivial and topological domains. The supercell is constructed with two symmetric interfaces: Interface 1 has the topological phase on the left and the regular phase on the right (in the \(+y\) direction), while Interface 2 reverses this arrangement. This configuration allows simultaneous evaluation of both pseudospin modes and their respective spatial localization.

\begin{figure} [h]
\centering
\includegraphics[width=0.7\textwidth]{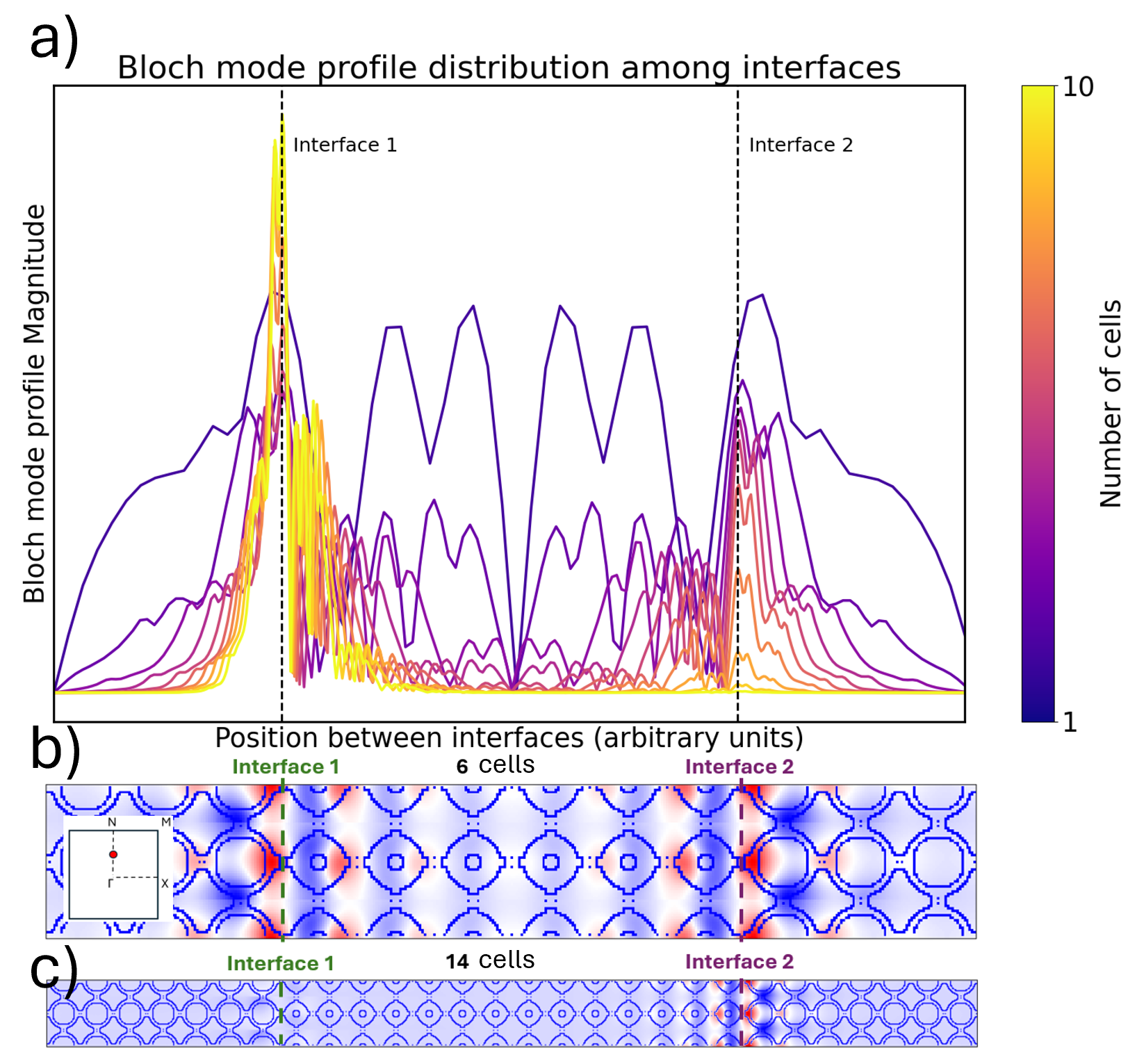}
\caption{\label{cell count}Bloch mode localization as a proxy for energy separation in individual pseudospin channels.  
\textbf{a)} Spatial magnitude of a topological Bloch mode along the \(x\)-axis, integrated over the \(y\)-direction of the supercell. Two interfaces are present: Interface 1 (topological phase on the left, regular on the right) and Interface 2 (reversed configuration).  
\textbf{b)} Field profile of a Bloch mode for a supercell composed of 6 cells of each type, evaluated at a \(k\)-point between \(\Gamma\) and \(N\). The mode is partially localized.  
\textbf{c)} Same scenario as in \textbf{b}, but with 14 cells of each type per supercell. A different pseudospin is shown, now localized around Interface 2, confirming that each topological mode spatially separates and associates with a distinct interface.
}
\end{figure}

Figure~\ref{cell count}\textbf{a} presents the spatial magnitude of the Bloch mode along the \(x\)-axis, integrated over the \(y\)-direction, for various supercell sizes. As the number of unit cells increases, the mode profiles become increasingly localized around the two interfaces, indicating that each pseudospin state isolates to its respective domain wall. This spatial confinement is a hallmark of topological protection and implies minimal mode mixing or backscattering under perturbations.

Figures~\ref{cell count}\textbf{b} and \textbf{c} provide further visualization of this localization. In Fig.~\ref{cell count}\textbf{b}, we show a representative Bloch mode in a supercell composed of six cells of each type. The mode is partially distributed across both interfaces. By increasing the number of cells to 14 (Fig.~\ref{cell count}\textbf{c}), we observe stronger confinement of each topological mode to a specific interface. Moreover, the figure illustrates that the two pseudospins localize independently, each around one interface, demonstrating the spatial separation of counter-propagating modes in a \(\mathbb{Z}_2\) topological system.

In summary, this analysis highlights the importance of including a sufficient number of PC periods adjacent to the interface. Doing so ensures not only spectral stability and strong pseudospin isolation, but also enables near-perfect unidirectional transport, a fundamental requirement for realizing robust topological photonic devices.

\section{Conclusion}

This work introduces a comprehensive methodology for the design and characterization of topological interfaces in two-dimensional photonic crystals, with a focus on realizing \(\mathbb{Z}_2\) topological phases. We address key challenges in symmetry tracking, band connection, and unit cell optimization, presenting a unified framework that links physical principles to computational tools for robust topological mode design.

Central to our approach is an iterative band connection algorithm that preserves mode identity across the Brillouin zone by relying on eigenvector symmetry, rather than energy ordering alone. This enables accurate classification of photonic bands, even near degeneracies and crossings. Complementing this, we introduce a symmetry recognition technique based on sign-processed eigenmodes and data-driven comparator generation, allowing consistent classification of modes across varying geometries without relying on predefined irreducible representations.

Using these tools, we mapped the parameter space of PC unit cells, identifying regions of partial band inversion essential for \(\mathbb{Z}_2\) interfaces. We demonstrated how geometric tuning controls eigenmode-permittivity overlap, enabling targeted band reordering and band-gap overlap. Our analysis of dispersion characteristics revealed trade-offs between band-gap size and group velocity, including configurations with anomalously flat bands for slow-light applications.

We further examined how the number of periods surrounding the interface influences the convergence, localization, and directionality of the topological modes. Simulations showed that with a moderate number of periods, the interface supports highly directional propagation with minimal backscattering. Supercell analyses revealed that counterpropagating modes localize on opposite sides of the structure, confirming robust pseudospin separation and topological protection.

Moreover, photonic crystals inherently permit scalability through the lattice parameter, allowing the operational wavelength to be shifted provided the chosen material remains in a low-absorption spectral range. This inherent scalability further enhances our framework’s adaptability for telecommunication frequencies and other targeted applications.

Altogether, this study offers a modular and scalable framework for engineering topological photonic interfaces, grounded in symmetry, geometry, and performance. The methodology is extendable to higher dimensions, dynamic materials, and machine learning–assisted design, paving the way for robust photonic platforms with applications in slow light, topological routing, and nonreciprocal devices.


\medskip
\textbf{Acknowledgements} \par 
We acknowledge financial support from Projects No. PID2023-152225NB-I00, and Severo Ochoa MATRANS42 (No. CEX2023-001263-S) of the Spanish Ministry of Science and Innovation (Grant No. MICIU/AEI/10.13039/501100011033 and FEDER, EU)"), Projects No. TED2021-129857B-I00 and PDC2023-145824-I00, funded by MCIN/AEI/10.13039/501100011033 and European Union NextGeneration EU/PRTR and by the Generalitat de Catalunya (2021 SGR 00445), We also acknowledge the financial support from the Charles University, project GA UK No. 80224 and the support provided by the Ferroic Multifunctionalities project, supported by the Ministry of Education, Youth, and Sports of the Czech Republic [Project No. CZ.02.01.01/00/22\_008/0004591]

\medskip
\textbf{Disclosures} \par 
The authors declare no conflicts of interest.

\medskip
\textbf{Ethics statement} \par
This study did not involve human participants, human data or tissue, or animal subjects.

\medskip
\textbf{Data Availability Statement} \par

The data that support the findings of this study are available from the corresponding author upon reasonable request.

\medskip

%
\bibliographystyle{MSP}
\bibliography{bibliography}

\end{document}